# TCSR-SQL: Towards Table Content-Aware Text-to-SQL With Self-Retrieval

Wenbo Xu[1] | Liang Yan[1,2] | Chuanyi Liu[1,3] | Peiyi Han[1,3] | Haifeng Zhu[1] | Yong Xu[1,3] | Yingwei Liang[4] | Bob Zhang[5]

[1]Department of Computer Science, Harbin Institute of Technology (Shenzhen), Shenzhen, China | [2]Inspur Cloud Information Technology Co. Ltd, Jinan, China | [3]Peng Cheng Laboratory, Shenzhen, China | [4]Guangdong Power Grid Co. Ltd, Guangzhou, China | [5]Faculty of Science and Technology, University of Macau, Taipa, China

**Correspondence:** Chuanyi Liu (liuchuanyi@hit.edu.cn)

## ABSTRACT

Large language model-based (LLM-based) text-to-SQL methods have achieved important progress in generating SQL queries for real-world applications. When confronted with table content-aware questions in real-world scenarios, ambiguous data content keywords and nonexistent database schema column names within the question lead to the poor performance of existing methods. To solve this problem, we propose a novel approach towards table content-aware text-to-SQL with self-retrieval (TCSR-SQL). It leverages LLM's in-context learning capability to extract data content keywords within the question and infer possible related database schema, which is used to generate Seed SQL to fuzz search databases. The search results are further used to confirm the encoding knowledge with the designed encoding knowledge table, including column names and exact stored content values used in the SQL. The encoding knowledge is sent to obtain the final *Precise SQL* following multi-rounds of generation-execution-revision process. To validate our approach, we introduce a table-content-aware, question-related benchmark dataset, containing 2115 question-SQL pairs. Comprehensive experiments conducted on this benchmark demonstrate the remarkable performance of TCSR-SQL, achieving an improvement of at least 27.8% in execution accuracy compared to other state-of-the-art methods.

## 1 | Introduction

Text-to-SQL (NL2SQL) is an academic research field that aims to transform natural language questions into corresponding structured query language (SQL) queries [1]. It can help people without programming skills to search data directly from databases relevant to the question, thus becoming a state of great interest and proliferation. Recently, the large language model (LLM) technique has rapidly developed and demonstrated its ability to emerge knowledge from massive amounts of data in various domains [2–4] and complete tasks related to data analysis [5, 6]. The latest text-to-SQL research focuses on designing LLM-based methods that use the database schema or data contents. Some of the representative methods employ solely database schema, including C3 [7], DIN-SQL [8], DAIL-SQL [9] and Fin-SQL [10]. Other methods employ both data contents and database schema, such as MAC-SQL [11], CodeS [12] and CHESS [13].

However, these studies still have difficulties in dealing with the problem of real-world scenarios: "How to generate SQL queries corresponding to table content-aware natural language questions?". As shown in Figure 1, there are two difficulties in solving this problem. First, the column names used by the gold SQL query generally do not appear in the data content keywords within the question. In the example, data content keywords appearing in the question ('the first quarter of 2023', 'GDP growth rate') are not the same as the column names used in the gold SQL query (‘reportperiod’, ‘indexname’, ‘cumulative’). Thus it is impossible for existing methods to directly use the data content keywords within the question as the column names, which may lead to errors in the generated SQL query. Second, the condition values used in the gold SQL query are not the same as the data content keywords within the question. In the example, the gold SQL query uses the condition (indexid = 5) to match the data content keyword ‘GDP growth rate’ within the question because it matches multiple values stored in the ‘indexname’ column (‘GDP growth(%)’, ‘GDP growth’, etc). Therefore, the gold SQL query solves this problem by utilising database-specific encoding knowledge with documented correlations (column ‘indexid = 5’ corresponds to column ‘indexname LIKE ‘%GDP growth%’’). However, the natural language question does not specify this correlation, making existing methods unable to either confirm the consistency of data content keywords within the question and content values stored in the database or apply the encoding knowledge. Thus incorrect condition values may be used when generating SQL queries.

Based on the preceding discussion, there are three technical challenges for LLM-based text-to-SQL methods to solve the above problem. First, it is arduous to input the entire database contents into LLMs. Real-world industrial databases usually store huge amounts of data. Due to the limited length of LLM input tokens, it will exceed the maximal input token length to embed all database contents and send them directly to LLM, making it computationally difficult to handle. Using LLMs with a larger input token length to complete the text-to-SQL task may result in high costs. Methods of adding data content samples [11, 12] cannot provide definitive results of the stored data content values used by SQL. Second, interference from other noisy data unrelated to the question tends to degrade the model performance. Real-world databases usually contain numerous data tables, rows and columns but only a small percentage are highly relevant to the question. Since the output result of LLMs will be directly affected by the input, directly inputting a large amount of irrelevant data into LLM may affect the LLM's understanding of the question and lead to errors in the output [14, 15]. Third, understanding all the data contents of databases is challenging. In practical database ecosystems, it is prevalent to find auxiliary data incorporated to facilitate maintenance tasks, such as an encoded knowledge table that meticulously maintains the alignment between domain-specific indexes and their corresponding encoded data columns. Inputting raw data from the database to LLM without prior processing may result in the loss of important prior knowledge, leading to errors in generating SQL queries that do not make use of such encoding knowledge. In summary, we need a more effective LLM-based text-to-SQL method that satisfies three aspects: the ability to handle large databases with massive amounts of data, robustness to noisy data and the ability to understand and utilise holistic database contents.

To address these technical challenges, we propose towards table content-aware text-to-SQL with self-retrieval (TCSR-SQL). The design of two self-retrieval techniques with fuzz testing and retrieval augmented generation (RAG) [16] enables TCSR-SQL to effectively obtain the database schema column names and exact stored content values, thus solving the above problem. Figure 2 illustrates the overall framework of TCSR-SQL, which comprises three modules: keywords extraction and fuzzy detection (Section 3.1), knowledge retrieval and alignment (Section 3.2) and SQL generation and revision (Section 3.3). The keywords extraction and fuzzy detection module extracts the data content keywords from the question, speculates the related database schema and searches the database for possible stored data content values. This module innovatively implements self-retrieval based on fuzz testing, thus eliminating the need for inputting the complete data contents. It generates a series of seed SQL queries to execute and identify any possible data contents in the specified columns. This helps estimate the exact content values stored in the database corresponding to the data content keyword within the question. The knowledge retrieval and alignment module serves to confirm the columns and values of the conditions used in the generated SQL queries based on the results of the first module. This module incorporates self-retrieval with the designed encoding knowledge table using RAG, thereby eliminating the impact of noisy data that is completely irrelevant to the question. Specifically, it incorporates both seed SQL's search results and the encoding knowledge table for condition confirmation. This significantly reduces the data scope to only database schema and data contents relevant to the question. The SQL generation and revision module generates a preliminary version of the SQL query (*Fuzzy SQL*) and iteratively modifies it to obtain the final result (*Precise SQL*). This module employs the findings of self-retrieval in the self-debugging process, thus enhancing the understanding of the database contents relevant to the question. The iterative generation-execution-revision process is achieved by using the SQL execution results and the encoding knowledge obtained in the second module. This facilitates an enhanced comprehension of the question, database schema and data contents.

| Natural Language Question | | Please answer what is the GDP growth rate for the first quarter of 2023 /请回答2023年一季度的GDP增速是多少 | | | | |
|---|---|---|---|---|---|---|
| **Database Schema** | **Table Name** | tj_qssczz_fcy | | | | |
| | **Column Names** | periodtype | periodyear | reportperiod /报告期 | indexname /指标名称 | indexid | cumulative /累计 |
| **Data Content Samples** | | quarter | 2023 | 2023Q1 /2023年一季度 | GDP /全市生产总值 | 1 | 234698.33 |
| | | quarter | 2023 | 2023Q3 /2023年三季度 | GDP(billion) /GDP（亿元） | 1 | 17.61 |
| | | quarter | 2023 | 2023Q1 /2023年一季度 | primary GDP /第一产业 | 2 | 5989.42 |
| | | quarter | 2023 | 2023Q4 /2023年四季度 | secondary GDP /第二产业 | 3 | 36233.54 |
| | | quarter | 2023 | 2023Q1 /2023年一季度 | GDP growth /生产总值增长 | 5 | 98.74 |
| | | quarter | 2023 | 2023Q2 /2023年二季度 | GDP growth(%) /GDP增长(%) | 5 | 36.20 |
| **Gold SQL Query** | | SELECT reportperiod, indexname, cumulative FROM tj_qssczz_fcy WHERE indexid = 5 AND reportperiod LIKE ‘%2023Q1%’ | | | | |

**FIGURE 1** | Problem example of the table content-aware question in real-world scenarios.

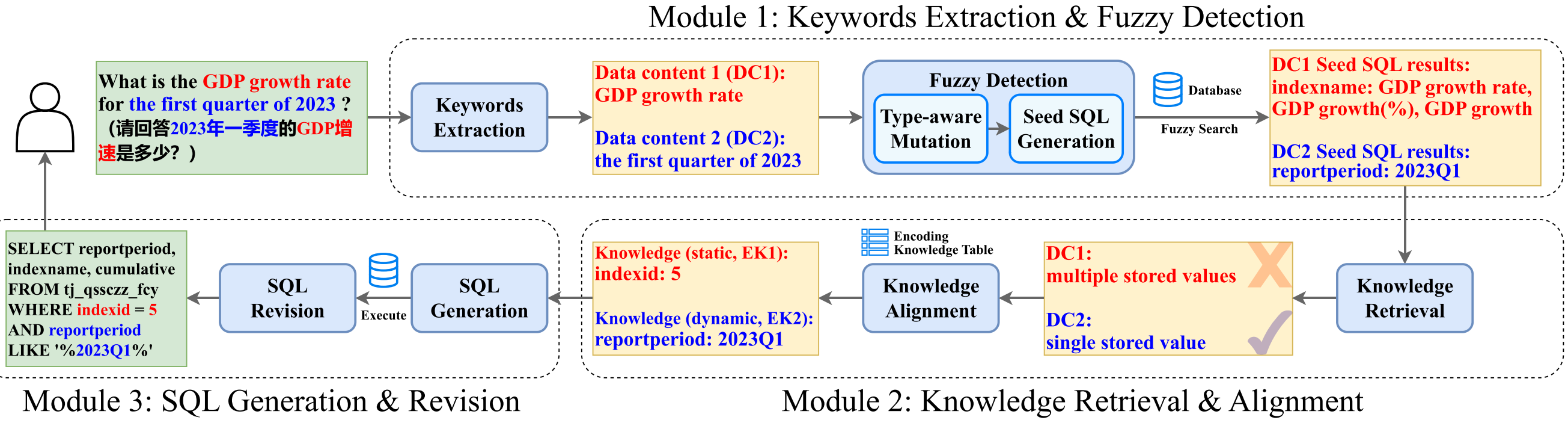

**FIGURE 2** | Overall framework of TCSR-SQL.

Existing general text-to-SQL benchmarks, such as Spider [17] and BIRD [18], are oriented towards validating methods against various problem types, with fewer table content-aware questions in real-world scenarios. In order to comprehensively evaluate our method against such questions, we create a table content-aware question-related benchmark test dataset (TCD) of 2115 question-SQL pairs. This dataset was derived from multiple Spider variant datasets and real-world databases. The execution and exact-set-match accuracy of our method reaches 77.8% and 53.5%, representing an improvement of at least 27.8% over existing LLM-based text-to-SQL SOTA methods. By analysing the comparison results in representative test cases, our method can identify the exact column names and stored content values used in SQL queries from the database. Through multiple rounds of generation-execution-revision interactions between LLM and the database, our method can further improve the generation accuracy of SQL queries.

Our contributions can be summarised as follows:

1. We propose TCSR-SQL with two innovative self-retrieval techniques based on fuzz testing and RAG, aiming at generating SQL queries corresponding to table content-aware questions from huge databases.
2. We create a novel benchmarking dataset called TCD, which includes 2115 question-SQL pairs to evaluate the effectiveness of solving the above problem.
3. We validate the effectiveness of our method on the constructed test dataset. Experimental results demonstrate that our method improves the performance by 27.8% compared to existing text-to-SQL SOTA methods.

In summary, TCSR-SQL aims to solve the problem of text-to-SQL for real-world table content-aware questions. Prior methods fail to deal with ambiguous data content keywords and nonexistent database schema column names within such questions. To address this, TCSR-SQL designs two novel self-retrieval techniques based on fuzz testing and RAG to effectively retrieve relevant column names and data content values stored in the databases.

## 2 | Related Work

Despite traditional text-to-SQL methods based on pretrained language models [19], the latest text-to-SQL studies pay close attention to utilising large language models (LLMs) for the text-to-SQL tasks. To surpass the performance of traditional methods, existing LLM-based text-to-SQL methods can be classified into two categories: database schema-based methods and data content-based methods.

### 2.1 | Database Schema-Based Text-to-SQL Methods

In the initial period of LLM-based text-to-SQL methods, research efforts primarily focus on using only database schema and developing basic NLP techniques to fully exploit the capabilities of LLMs for this task. According to the type of techniques, these methods can be categorised into two groups.

*Prompt-based methods*. The first representative technique is prompt engineering, including least-to-most prompting, zero-shot prompting, few-shot/compound prompting, chain-of-thought (CoT) and self-debugging. For example, C3 [7] designs a zero-shot prompting technique by bifurcating the text-to-SQL task into two sub-tasks: schema linking and SQL generation. DIN-SQL [8] designs a multi-round few-shot prompting strategy by decomposing the text-to-SQL task into four subtasks: schema linking, classification and decomposition, SQL generation and self-correction. It integrates the least-to-most prompting, chain-of-thought (CoT) prompting and self-debugging techniques into the few-shot prompting technique, thereby elevating the text-to-SQL performance of LLMs. DAIL-SQL [9] analyses and integrates the advantages of existing prompt-based text-to-SQL methods from three perspectives: question representation, example selection and example organisation. It chooses the code representation prompt (CR) in the question representation stage, masked question similarity selection (MQS) and query similarity selection (QRS) in the example selection stage and SQL-only organisation (SO) in the example organisation stage. ACT-SQL [20] provides a method to automatically generate auto-CoT text-to-SQL exemplars, thereby eliminating the need for manual labelling. Zhai et al. [21] employ CoT to provide guidance as prompts through learning from mistakes in the training data. On the premise of providing database schema only, some recent studies focus on combining various prompt techniques for specific problems in SQL generation [22–24].

However, these methods face several limitations. Limited by the length of LLM input tokens and the database schema-only input, it is difficult to retrieve the full contents of a large real-world database. This leads to errors in the database storage values used in the conditions of the generated SQL as well as incorrect column names used in the generated SQL.

*Fine tuning-based methods.* With the rapid iteration of LLM techniques and the large-scale emergence of open-source LLM models, there are several studies devoted to training general LLMs into text-to-SQL domain-specific LLMs by using fine-tuning techniques. As an initial attempt, SQL-PALM [25] and DAIL-SQL [9] utilise the Spider training set [17] to discover the fine-tuning effectiveness of the closed-source PaLM-2 [26] and several open-sourced LLMs, separately. Fin-SQL [10] uses LoRA [27], a common technique for LLM fine-tuning, to design the LoRA-based multi-task PEFT method. It fine-tunes different open-source LLMs to complete the text-to-SQL task for financial domain databases, including LLaMA-2 [28] and Baichuan-2 [29]. DTS-SQL [30] proposes a two-stage fine-tuning approach for data privacy protection issues using open-source LLMs with a small number of parameters. By decoupling the task into schema linking fine-tuning and SQL generation fine-tuning, it improves the performance of fine-tuned LLMs to a level close to existing prompt-based text-to-SQL methods using closed-source LLMs such as GPT-4.

These fine tuning-based methods have three limitations. First, most of them use the training dataset of the specified benchmark to fine-tune LLMs. This results in poor robustness of fine-tuned LLM when migrating to text-to-SQL tasks with other irrelevant data. Second, the text-to-SQL domain-specific fine tuning process implicitly teaches LLM to capture the correlations between fine tuning data, questions and SQL queries. The accuracy might be significantly reduced when dealing with issues that necessitate explicit database-associated encoding knowledge. Third, these methods' inputs still use only database schema, which may lead to incorrect condition values used in the SQL.

## 2.2 | Data Content-Based Text-to-SQL Methods

In real-world applications, researchers have identified that manually writing SQL queries is likely to result in invocation and viewing of the data content within the database. Therefore, recent LLM-based text-to-SQL methods also consider adding data contents to further improve the task performance.

*LLM-only methods.* Some of them directly add several data content samples based on the prompt-based and fine tuning-based methods mentioned above. For prompt-based methods, SQLPrompt [31] adds relevant database content values in the input prompts based on SQL-PALM [25]. DFIN-SQL [32] enhances the schema linking module of the original DIN-SQL [8] by adding sample rows with data contents. PET-SQL [33] utilises data content samples in its schema linking and finSQL generation prompt. It further introduces diverse LLMs to facilitate the cross consistency and fine-grained voting. MCS-SQL [34] introduces data content samples to facilitate the generation of multiple SQL queries and the subsequent selection of the desired result with high temperature. For fine tuning-based methods, CodeS [12] incorporates data content samples into the designed database prompt construction session to fine-tune the text-to-SQL domain-specific LLM. SQLCritic [35] leverages CodeS to design a training strategy, which is used to obtain the SQL error-checking LLM.

However, there is a significant problem with the above methods. The data contents are combined with other related information within the prompt. It is uncertain whether LLM understands and applies the data contents since there is no feedback. Furthermore, it is unclear how the data contents can be used to help improve the accuracy of SQL generation.

*Agent-based methods.* Recently, the LLM-based agent technique has attracted considerable attention due to its ability to introduce additional information to significantly enhance the performance of a single LLM across various tasks. Several studies in the text-to-SQL domain explore the possibility of agent-based text-to-SQL methods. For example, Guo et al. [36] and SQL-CRAFT [37] both incorporate feedback on the execution of SQL queries from the database into the prompt. They design a dynamic revision chain and an automated loop control algorithm to guide LLMs to refine their generated SQL queries, separately. Knowledge-to-SQL [38] and Dubo-SQL [39] generate SQL queries based on the idea of retrieval augmented generation (RAG) [16]. The former introduces a trained data expert LLM (DELLM), which is trained using the gold knowledge in the BIRD benchmark [40]. The DELLM further generates additional knowledge to help LLMs improve SQL accuracy. The latter claims to introduce diverse RAG by comparing vector embeddings of questions and selecting few-shot examples from the BIRD training set. MAC-SQL [11] designs three LLM-based agents (i.e., selector, decomposer and refiner) and uses the multi-agent collaboration framework to address challenges in text-to-SQL tasks. It provides a foundation for further research in this area [41, 42]. CHESS [13] creates a vector database to repeatedly invoke the embedding vectors for each evidence in the BIRD benchmark [18] during its designed three modules: entity and context retrieval, schema selection and query generation.

Compared to the usage of data content, such as the revision loop in SQL-CRAFT, the selector agent in MAC-SQL and the evidence invocation in CHESS, the uniqueness of TCSR-SQL lies in the two novel self-retrieval techniques based on fuzz testing and RAG. The former introduces type-aware mutation seed pools used to generate seed SQLs to search for potential column names and stored data contents relevant to the keywords within the question. The latter leverages the above search results and the designed encoding knowledge table to confirm the columns and condition values used in the generated SQL queries. Therefore, TCSR-SQL can address ambiguous data content keywords and nonexistent database schema column names in table content-aware questions.

# 3 | TCSR-SQL

Existing LLM-based text-to-SQL methods face two difficulties in accurately generating SQL queries according to table content-aware natural language questions from a large database:

identifying column names and exact content values stored in the database. In this paper, we propose TCSR-SQL to overcome these difficulties. The logic behind TCSR-SQL lies in leveraging the in-context learning capabilities of large language models (LLMs) to extract relevant database information from questions. By interacting with the database, TCSR-SQL incorporates accurate column names and exact content values into the prompt, thus enabling the LLM to self-retrieve database content for text-to-SQL. As shown in Figure 2, TCSR-SQL consists of three modules:

1. *Keywords Extraction and Fuzzy Detection*. This module implements the self-retrieval technique with the idea of fuzz testing. It takes a natural language question, database schema and database content samples as input. It then extracts the data content keywords from the question and all the possible related columns in the database. They are further sent to the designed fuzzy detection session to search for potentially relevant content values stored in the database.
2. *Knowledge Retrieval and Alignment*. This module implements the self-retrieval technique based on the idea of retrieval augmented generation (RAG) [16]. It takes the output of the first module (i.e., data content keywords and search results for potentially relevant content values) as input. It then tries to identify the exact stored content values corresponding to the data content keywords according to the search results. For unascertainable situations, it matches the alternative column names and stored content values using the designed encoding knowledge table. The final output is the encoding knowledge for each data content keyword, including both the column names and exact content values, stored in the database.
3. *SQL Generation and Revision*. This module takes the original input from the first module and the encoding knowledge from the second module as input. It then generates the *Fuzzy SQL* based on the question, database schema and database content samples. The *Fuzzy SQL* is further modified based on the database column names and exact stored content values provided in the encoding knowledge. The output is the *Precise SQL*, achieved by executing the SQL and revising incorrect *Fuzzy SQL* according to the execution feedback.

## 3.1 | Keywords Extraction & Fuzzy Detection

Algorithm 1 shows the workflow of the keywords extraction and fuzzy detection module and Figure 3 presents an example. The first objective of this module is to identify the database schema used in the generated SQL query based on the inaccurate data content keywords extracted from the natural language question. Another objective is to determine the content values stored in the database associated with the data content keywords as much as possible. The technical challenges lie in two aspects. First, it is impossible to input the complete table contents of a large database into large language models (LLMs) due to the upper limit of the input context token length of LLMs. Second, generating SQL queries for natural language questions only requires a small proportion of the data stored in the database, resulting in interference from other noisy data unrelated to the question.

### 3.1.1 | Keywords Extraction

First, we design a few-shot prompt-based method for keyword extraction. The input prompt consists of a question, database schema and four randomly selected data samples for each table in the database. LLM is asked to extract the data content keywords within the question, leveraging its in-context semantic learning capability. Based on extracted data content keywords and provided table samples, LLM further confirms the required database schema (i.e., table names and column names). It should be emphasised that we require LLM to identify a unique table name and several associated column names. The purpose of this approach is to maximise the probability of finding the stored content values associated with the inaccurate data content keyword within the question, while accurately assigning the keyword to the related data table at the same time. It allows for the verification of the column names and content values stored in the database that are required to generate the final SQL query. *DC*1 and *DC*2 in Figure 3 illustrate two examples, and the full prompt can be found in Figure 4.

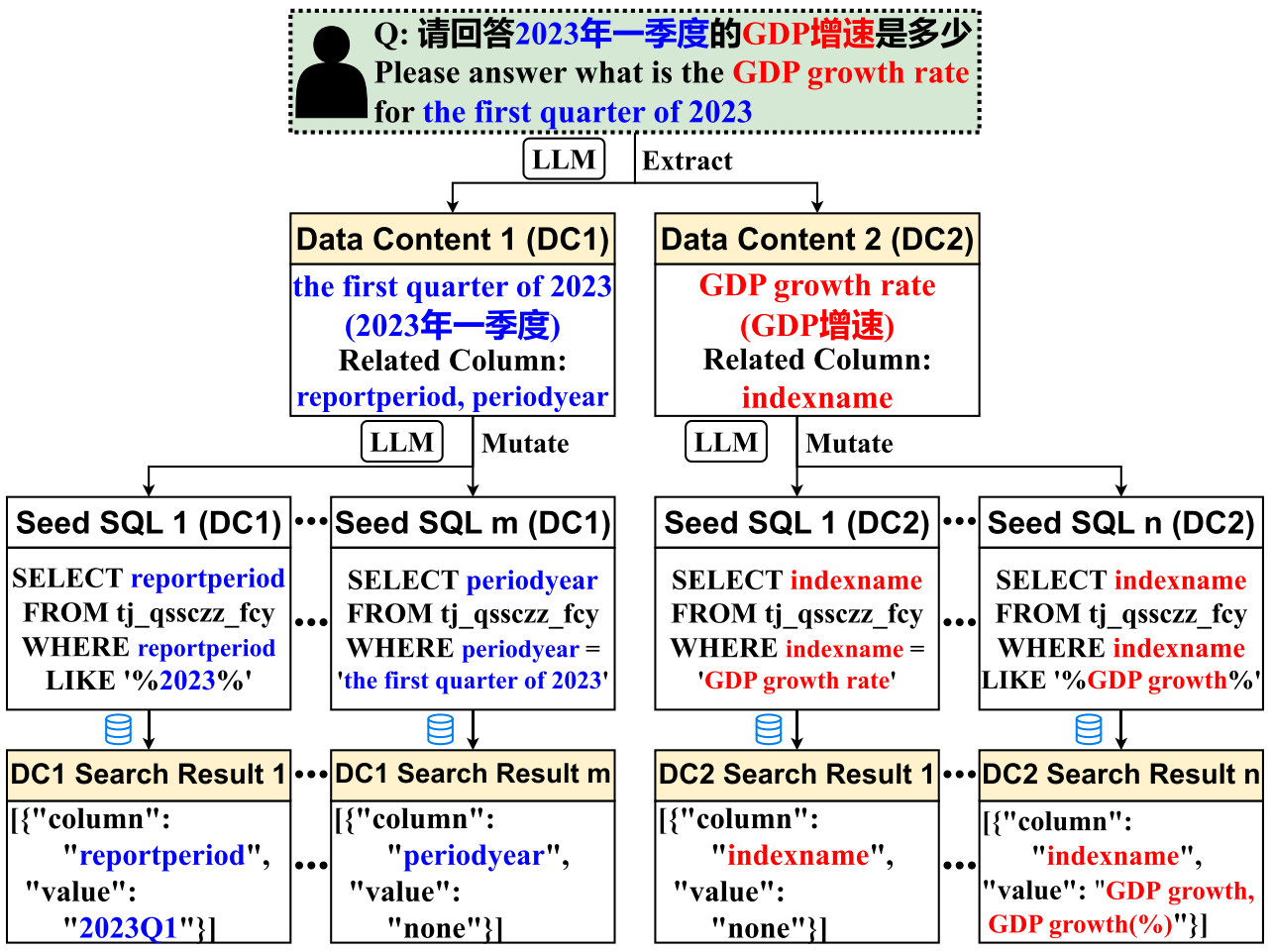

**FIGURE 3** | Example of the keywords extraction and fuzzy detection.

```
### First, find schema keywords and data content
keywords in the given question. Then, for each keywor
try to identify the possible corresponding tables and
columns using the value examples provided.

==========
{4 demonstrations}
==========

### Sqlite SQL tables, with their table names, column
names and data value examples:
{desc_str}
### Foreign keys of SQLite tables, used for table
joins:
{fk_str}
### Question:
{query}
### Answer:
```

**FIGURE 4** | Keywords extraction prompt template.

**ALGORITHM 1** | Keywords extraction & fuzzy detection.

**Input:**
Database schema, $DS$;
Natural language question, $Q$;
Data content samples, $Sam$.
**Output:**
Search results of each Seed SQL, $SR$.

```
for each Q_i in Q do
    DC, CN = Extract(DS, Q_i | LLM);
    // data content DC and column name CN
    for each DC_j in DC do
        DC ← Mutate(DC_j, Sam | LLM);
    SeedSQL = Generate(DC, CN, Skeleton);
    for each SeedSQL do
        SR ← Execute(SeedSQL);
return SR;
```

### 3.1.2 | Fuzzy Detection

Based on the data content keywords extracted from the question and the corresponding database table names and column names, we design a few-shot prompt-based method for fuzzy detection by innovatively combining dynamic fuzz testing with the text-to-SQL task. For each data content keyword, the fuzzy detection session needs to generate a series of SQL queries (i.e., seed SQL) to execute and search for relevant data content values in the specified columns. These search results are sent to the subsequent module to identify the exact stored content value corresponding to the inaccurate data content in the question. As shown in Figure 3, we introduce LLM to generate a series of diverse seed SQL to facilitate this session in accordance with a typical mutation-based fuzzing workflow [43]:.

1. *Type-aware mutation*. Inspired by Liu et al. [44], we divide the seed SQL into three types of mutually independent components (i.e., seed type) in this process. For each component (i.e., seed), we initialise the seed pool and bootstrap the seed SQL generation pipeline. As shown in Table 1, a single seed SQL has three types of mutable components: column names, data contents and SQL skeletons. For column names, we set all possible related database column names obtained in the previous keywords extraction session as the seed pool. For data contents, we first introduce LLM to predict the synonyms of the data content keywords within the original question. As shown in Figure 5, the input prompt includes data content keywords and data samples from related columns. Then we set the synonyms suggested by LLM and the corresponding data content keyword as the seed pool. For SQL skeletons, we set the following two candidate SQL query templates as the seed pool:

SQL1: **SELECT DISTINCT** {CN} **FROM** {TN}
**WHERE** {CN} = {DC};

SQL2: **SELECT DISTINCT** {CN} **FROM** {TN}
**WHERE** {CN} **LIKE** '%{DC}%'.

2. *Seed SQL generation and execution*. For each data content keyword $DC_i$, we traverse each seed pool of column names, data contents and SQL skeletons, respectively, obtaining a different set each time. Each set is used to construct a new seed SQL (i.e., seed SQL $j$ for $DC_i$ in Figure 3) to execute and obtain the results (i.e., $DC_i$ search result $j$ in Figure 3). In this way, we ensure that all seed SQLs are distinct and executable while maximising the search for data content values related to $DC_i$.

**TABLE 1** | List of type-aware mutation over Seed SQL.

| Component type | Mutation |
|---|---|
| Column name ($CN$) | *Return*: a random $CN$ (selected from related columns) |
| Data content ($DC$) | *Return*: a synonym related to $DC$ (generated by LLM) |
| SQL skeleton | *Return*: a random SQL skeleton (selected from SQL template candidates) |

## 3.2 | Knowledge Retrieval & Alignment

Algorithm 2 shows the workflow of the knowledge retrieval and alignment module and Figure 6 presents an example. The objective of this module is to identify the exact content value stored in the database by using the database's value search results of related data content keywords provided in the above module. The technical challenge is that it is difficult for LLMs to fully comprehend all the contents of the database and the database-specific encoding knowledge, leading to multiple errors when generating the final SQL query.

```
### Based on data samples extracted from the related
columns, infer the possible storage values (up to 5) of
the data content keyword in the database. Please
provide them in the form of a list.

{9 demonstrations}

# data content keyword: {keyword}.
# data samples for column {column}: {datasamples}
# possible storage values:
```

**FIGURE 5** | Type-aware mutation prompt template for $DC$.

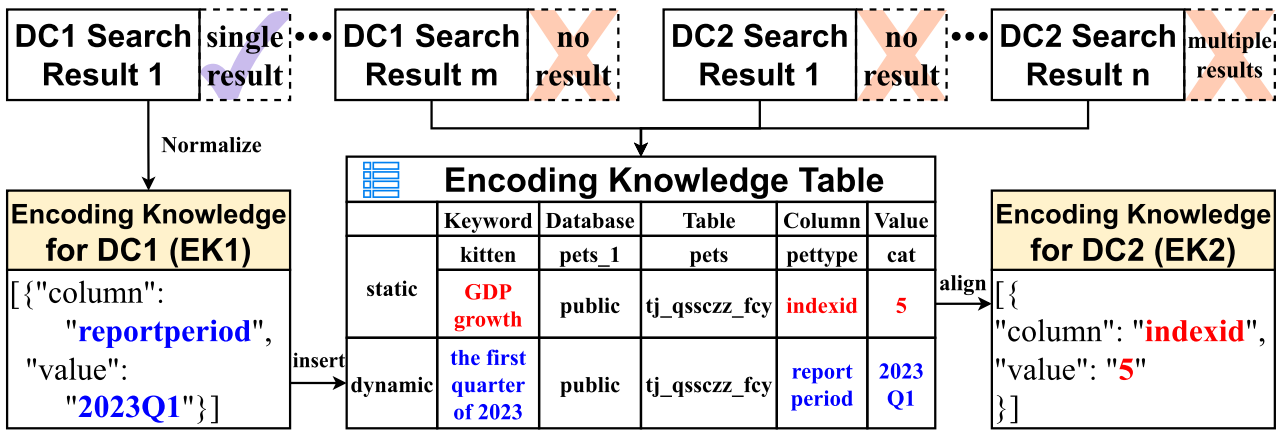

**FIGURE 6** | Example of the knowledge retrieval and alignment.

### 3.2.1 | Knowledge Retrieval

First of all, we start the knowledge retrieval session by categorising the search results from the output of the fuzzy detection session:

1. *Case* 1. The search results indicate that there is a unique content value exactly stored in the associated column corresponding to the data content keyword within the question. We normalise the search results to encoding knowledge and add it into the encoding knowledge table as available candidate knowledge for generating the final SQL query. Here, the normalisation process involves the data content keyword within the original question, the associated database name, the associated table name, the associated column name and the exact stored content value in JSON format. The normalisation result is then merged into the encoding knowledge table, which corresponds to the data content keywords within the question.
2. *Case* 2. The search results indicate that there are multiple content values stored in the associated column corresponding to the data content keyword within the question or the exact content value cannot be found in the currently associated column. Therefore, we consider the current search results invalid and discard them directly.

**ALGORITHM 2** | Knowledge retrieval & alignment.

**Input:**
Data content keywords in the question, $DC$;
Search results of each Seed SQL $SR$;
Encoding knowledge table, $T_{EK}$.
**Output:**
Encoding knowledge for data content keywords, $EK$.

**for** each $DC_i$ in $DC$ **do**
  **for** each $SR_j$ in $SR$ **do**
    **switch** the data content store value of $SR_j$ **do**
      **case** a unique value **do**
        $T_{EK} \leftarrow Normalization(SR_j)$;
      **case** multiple values **or** no value **do**
        remove $SR_j$ from $SR$;
  $Emb_{DC_i} = Embedding(DC_i)$;
  **for** each $EK_{i'}$ in $T_{EK}$ **do**
    $Emb_{EK_{i'}} = Embedding(EK_{i'})$;
    $Sim^{DC_i}_{EK_{i'}} = cosine(Emb_{DC_i}, Emb_{EK_{i'}})$;
  $EK \leftarrow EK_{i'}$ with $max(Sim^{DC_i}_{EK_{i'}})$;
**return** $EK$;

It should be noted that the encoding knowledge table mentioned above is designed based on the idea of retrieval augmented generation (RAG) [16]. Specifically, it consists of five columns: the data content keyword within the question, the associated database name, the associated table name, the associated column name and the exact stored content value. The construction of the encoding knowledge table ($T_{EK}$) involves static and dynamic knowledge. The static knowledge requires manual collection from the relationship-matching table in the database, which records the correspondence between the domain synonyms and specific encoding columns stored in the database. For example, $EK_1$ in Figure 6 is the static knowledge of the keyword 'GDP growth rate' within the question. This keyword has multiple synonyms stored in the associated column and they all correspond to the value '5' of the 'indexid' column specified in the relationship-matching table. Therefore, we manually construct the encoding knowledge as follows: {data content: 'GDP growth', database: 'public', table: 'tj_qssczz_fcy', column: 'indexid' and value: '5'}. The dynamic knowledge is derived from the above *Case 1* automatically. For example, $EK_2$ in Figure 6 is the dynamic knowledge of the keyword 'the first quarter of 2023'. This keyword corresponds to a unique content value stored in the associated column. Thus, we automatically normalise the value as a new candidate knowledge: {data content: 'the first quarter of 2023', database: 'public', table: 'tj_qssczz_fcy', column: 'reportperiod' and value: '2023Q1'}. Introducing two types of knowledge in the encoding knowledge table has two advantages. First, it contains the static knowledge used in the database construction process, which helps LLM acquire prior information about the database. Second, continuous updating of dynamic knowledge can help LLM memorise the exact data content of unseen or dynamically evolving databases, thus reducing operational costs of static knowledge and further improving the accuracy of SQL queries. Removing either knowledge can cause performance degradation. For example, the removal of static knowledge results in a 21.3% drop in execution accuracy and a 16.1% drop in exact-set-match accuracy, respectively. However, our method's performance is still competitive with only dynamic knowledge, which will be further discussed in Section 4.3.

### 3.2.2 | Knowledge Alignment

Based on the updated encoding knowledge table output from the knowledge retrieval session, we develop the knowledge alignment session aiming at further understanding the database content related to the inaccurate data content keywords within the natural language question. Specifically, we use the embedding model to compute the embedding representation of the data content keywords within the question and each piece of encoding knowledge:

$$
\begin{aligned}
Emb_{DC_i} &= Embedding(DC_i),\\
Emb_{EK_{i'}} &= Embedding(EK_{i'}),
\end{aligned} \tag{1}
$$

where $DC_i$ denotes the $i$-th data content keyword extracted from the question, and $EK_{i'}$ denotes the $i'$-th knowledge in the encoding knowledge table. Next, we calculate the cosine similarity between the embedding of the data content keyword and the embedding of the encoding knowledge:

$$
Sim^{DC_i}_{EK_{i'}} = cosine(Emb_{DC_i}, Emb_{EK_{i'}}). \tag{2}
$$

Then, we choose the encoding knowledge with the highest cosine similarity score for each data content keyword within the question. The knowledge contains the column name from the database schema and the exact content value stored in the database, which can be used as the available knowledge to generate the SQL query. Finally, all of the chosen encoding knowledge is sent to the subsequent module to modify the *Fuzzy SQL* and produce the final *Precise SQL*.

### 3.3 | SQL Generation & Revision

In the knowledge retrieval and alignment module, we obtain encoding knowledge including column names and exact content values stored in the database corresponding to the data content keywords, which could be used in the generation of the final SQL query. Furthermore, the inclusion of the self-correction techniques [45] is essential for checking and correcting the generated answers, especially in the context of text-to-SQL tasks [8, 11]. Based on the generation-execution-revision framework, we combine the obtained encoding knowledge and existing self-correction techniques and propose the SQL generation and revision module.

As shown in Figure 7, the objective of this module is to generate SQL queries using encoding knowledge and automatically rectify them based on the error feedback. First, we prompt LLM to generate a preliminary SQL query called *Fuzzy SQL* based on the natural language question, the database schema and the database content samples. We call it *Fuzzy SQL* because this SQL query is generated by LLM after fully understanding the question based on the input prompt. Although the *Fuzzy SQL* has a complete SQL skeleton, the column names and used data contents are directly derived from the original question, which is uncertain and might be modified. Next, we introduce LLM to revise the *Fuzzy SQL* using the encoding knowledge. The encoding knowledge includes column names and precise content values corresponding to the data content keywords within the question. The revised *Fuzzy SQL* is then executed to evaluate its correctness, feasibility and ability to retrieve nonempty results from the database. If the evaluation is approved, the result will be output as the final *Precise SQL*. Otherwise, the result will be further revised and re-evaluated. The above process is repeated until the result is correct or the number of revisions reaches the upper limit. The full prompt can be found in Figures 8 and 9.

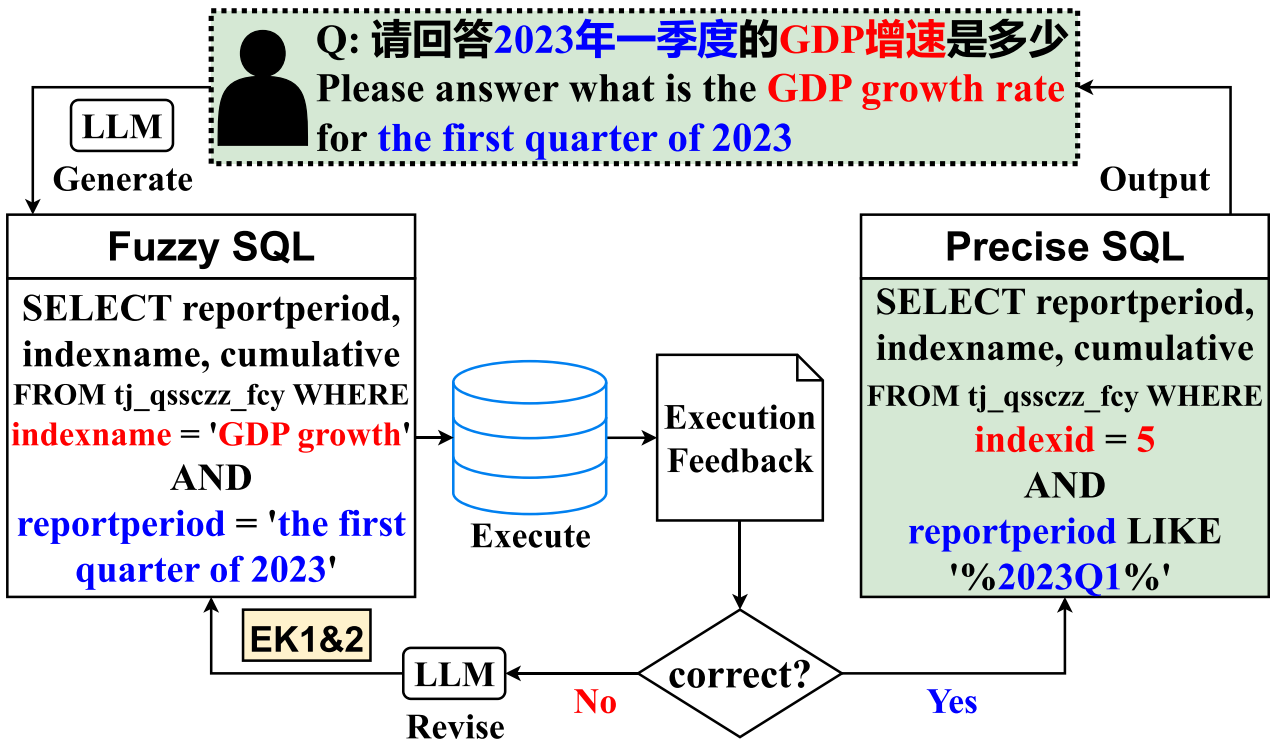

**FIGURE 7** | Example of the SQL generation and revision.

## 4 | Evaluation

### 4.1 | Experiment Setup

#### 4.1.1 | Datasets

Existing general text-to-SQL benchmarks, such as Spider [17] and BIRD [18], are oriented towards validating methods against various problem types, with fewer table content-aware questions in real-world scenarios. To evaluate the model performance on

```
### Refer to the provided table and encoding knowledge,
follow the requirements, and use valid SQLite to answer
the question.

### Requirements:
# In `SELECT <column>`, just select needed columns in
the Question without any unnecessary column or value
# If the same column name appears in multiple tables,
specify the table to which it belongs to avoid
ambiguity.

==========
{4 demostrations}
==========

### Sqlite SQL tables, with their table names, column
names and data value examples:
{desc_str}
### Foreign keys of SQLite tables, used for table
joins:
{fk_str}
### Question:
{query}
###  Encoding knowledge for SQL generation:
{related_prompt}
### SQL:
SELECT
```

**FIGURE 8** | SQL generation prompt template.

```
### When executing SQL below, some errors occurred,
please fix up SQL based on query and database info.
### Solve the task step by step if you need to. Using
SQL format in the code block, and indicate script type
in the code block.
### When you find an answer, verify the answer
carefully.

### Notes & Examples:
# In `SELECT <column>`, just select needed columns in
the Question without any unnecessary column or value
# Don't use `IN`, `OR`, `LEFT JOIN` as it might cause
extra results, use `JOIN`, `INTERSECT`, `EXCEPT`
instead
# Use `DISTINCT`, `DESC`, or `LIMIT` when necessary
# In `FROM <table>` or `JOIN <table>`, do not include
unnecessary table
# In `JOIN <table>`, make sure the Foreign keys used in
the SQL is correct

### Question:
{query}
### Encoding knowledge for SQL generation:
{related_prompt}
### Sqlite SQL tables, with their table names, column
names and data value examples:
{desc_str}
### Foreign keys of SQLite tables, used for table
joins:
{fk_str}
### Old SQL:
{old_sql}
### SQLite error:
{sqlite_error}
### Exception class:
{exception_class}

### Now please fixup old SQL and generate new SQL only
and with no explanation.
### correct SQL:
```

**FIGURE 9** | SQL revision prompt template.

this type of question, we create a table content-aware question-related benchmark test dataset (TCD) with 2115 question-SQL (QA) pairs. As shown in Table 2 [10, 46–49], we select 1269 table content-aware QA pairs from four Spider-variant datasets [46–49] by removing those with identical problems, empty execution results or semantic errors. We further add 423 QA pairs by rewriting relevant questions from a financial dataset called BULL [10], and manually annotate 423 QA pairs based on two real-world business databases.

*Spider-DK* [46] modifies QA pairs in the Spider-dev set by adding domain knowledge that reflects the interpretation of real-world difficulties. They consider five types of questions involving domain knowledge, three of which are related to the table content-aware question: T3 requires the model to recognise cell value synonym substitution. T4 requires the model to understand conditions for encoding corresponding to the question. T5 requires the model to understand the ambiguous column names within the question.

*Spider-Syn* [47] modifies natural language (NL) questions in the Spider benchmark by replacing column names and cell values with manually selected synonyms. We choose table content-aware QA pairs by selecting questions without the exact database schema or cell values.

*Spider-Realistic* [48] manually modifies a subset of Spider-dev questions, including columns, compared against values or used in clauses such as ORDER BY. We select table content-aware QA pairs by choosing those questions without explicit mention of column names.

*Dr.Spider* [49] designs 17 datasets after perturbation of databases, NL questions and SQL queries to measure robustness from different perspectives. Two of them are related to the table content-aware question. In column-value, column names used in SQL need to be inferred from values mentioned in the question. In value-synonym, values mentioned in the question are replaced with synonyms.

*BULL* [10] is collected from the financial analysis company, including large practical databases for fund, stock and macro economy. We obtain table content-aware QA pairs by rewriting several Chinese questions based on the fund and stock databases.

**TABLE 2** | Construction details of TCD.

| Name | Source | Type | Sum of QA pairs | Data size |
|---|---|---|---|---|
| TCD | DK [46] | Cross-domain | 107 | 8596 |
| | Syn [47] | | 254 | |
| | Realistic [48] | | 210 | |
| | Dr. [49] | | 698 | |
| | BULL [10] | Finance business | 423 | 220,458 |
| | Real-world | | 423 | 14,341 |

#### 4.1.2 | Evaluation Metrics

*Execution accuracy (EX)* compares the execution output of the predicted SQL query to that of the gold SQL query on a question. Since there may be multiple valid SQL queries for a single question, the EX metric can match SQL queries that execute with the same result regardless of differences in the string representation between the generated SQL and the gold SQL.

*Exact-set-match accuracy (EM)* compares each clause in the predicted SQL to that in the gold SQL. A predicted SQL is considered if all of its components match the ground truth. Although not meet the requirements of exact-set-match, some predicted SQL can still be executed to produce correct results. Therefore, the EM metric is only used as a reference result.

#### 4.1.3 | Baselines

We conduct experiments on the TCD test dataset and compare TCSR-SQL with representative baselines of database schema-based and data content-based methods:

C3-SQL [7] is a zero-shot text-to-SQL method with three components, which correspond to the model input, model bias and model output, respectively. DIN-SQL [8] decomposes text-to-SQL tasks into smaller sub-tasks and designs different prompts for each sub-task to instruct GPT to obtain the final SQL. DAIL-SQL [9] selects few-shot demonstrations based on their skeleton similarities and removes cross-domain knowledge from examples for token efficiency. MAC-SQL [11] is a novel LLM-based multi-agent collaborative framework for text-to-SQL consisting of a selector, a decomposer and a refiner. CodeS [12] is a series of open-source fine-tuned LLMs tailored for text-to-SQL tasks. We choose CodeS-15b-spider as the baseline because it is the largest model fine-tuned on the Spider-train set, which is the same as fine-tuning on the Spider-variant datasets of our TCD dataset. CHESS [13] employs three sessions with preprocessing acceleration techniques and outperforms all other open-source methods on the BIRD benchmark.

#### 4.1.4 | Settings

We use GPT-3.5-turbo [50] in our experiments and the temperature is set to 0. The reason for solely using GPT-3.5-Turbo is to achieve competitive performance using a general-purpose LLM with moderate capabilities, demonstrating the unique effectiveness of TCSR-SQL rather than the advanced ability of the latest LLMs. We choose the Dmeta-embedding algorithm[1] as the embedding model for similarity calculation in the knowledge retrieval and alignment module. Table 3 shows the configuration of our experiments.

### 4.2 | Overall Results

Table 4 shows the main results of TCSR-SQL compared with all baseline methods. The best are marked in bold and the second-best are underlined. Compared to the second-best baseline, TCSR-SQL improves EX by at least 27.8%, which reaches 77.8% EX and 53.5% EM. For each of the generated SQL queries, TCSR-SQL achieves 607 token cost, 113.28 SQL length and 0.36 number of joins on average. C3-SQL is poor in inferring conditional values due to the lack of checking of data contents. Thus, the EX accuracy performs the worst (30.1%) and the EM accuracy is only 22.4%. The EX results of DIN-SQL and DAIL-SQL based on few-shot prompting are about 33%. Although DAIL-SQL selects more suitable examples from the Spider-train set for different questions, and DIN-SQL designs the prompts carefully with self-correction, their examples and the correction mechanism are not based on execution results. However, the execution results happen to be the direct reference for correctness. MAC-SQL utilises database values to simplify the database schema, decomposes the question and uses execution results for self-correction. Although the EM performs the worst, the EX accuracy reaches 44.2%. This indicates that it has a good semantic understanding of the question, although it does not adopt a sentence structure similar to the gold SQL. CodeS-15b-spider achieves the second highest EX and EM with the least token consumption because it utilises four Spider-variants of our TCD dataset for time-consuming fine-tuning to generate similar SQL skeletons. Despite the good inference ability by adding the most relevant values extracted from the data content in the prompt, the EX accuracy remains 27.8% lower than TCSR-SQL. CHESS exhibits suboptimal performance, possibly due to its inability to construct and capture valid supplementary information by directly leveraging evidence from BIRD during the preprocessing phase. TCSR-SQL performs better because of the self-retrieval techniques to help obtain the correct columns and conditional values used in the generated SQL queries.

**TABLE 3** | Parameter configuration.

| Parameters | Values |
|---|---|
| GPT model API | Gpt-3.5-turbo-0125 |
| Temperature | 0 |
| Number of data content samples | 6 |
| Examples of keywords extraction | 4-Shot |
| Examples of fuzzy detection | 9-Shot |
| Examples of SQL generation | 4-Shot |
| Examples of SQL revision | 0-Shot |

## 4.3 | Ablation Experiments

In this section, we evaluate the impact of the existence of each module as shown in Table 5.

*w/o Keywords Extraction and Fuzzy Detection.* As no keywords are extracted, we directly use the question to search the encoding knowledge table and add the first three pieces of static knowledge to the prompt as references.

The keywords extraction and fuzzy detection module can find related tables and columns through keyword extraction. Without this module, the EX drops by 3.3%, whereas the EM rises a little. In this module, LLM performs keyword extraction and preliminarily confirms whether it is related to the database schema or data values, thereby further detecting keywords related to data values in the corresponding columns. The keywords related to data contents are often values in SQL conditions, whereas keywords related to database schema are often columns in SQL conditions or used for selection. Incorrect data values or columns will directly lead to the generation of incorrect SQL queries. Without this module, the keywords within the question cannot be explicitly linked and further explored or searched for knowledge, resulting in the inability to delve into the database directly to check the columns and exact stored values.

*w/o Knowledge Retrieval and Alignment.* We only use the dynamic knowledge from the first module to help generate SQL and skip the knowledge retrieval and alignment module.

The knowledge retrieval and alignment module finds the encoded representation of relevant semantics in the database by

**TABLE 5** | Ablation results.

| Methods | EX | EM |
|---|---|---|
| **TCSR-SQL** | **77.8** | **53.5** |
| w/o keywords extraction and fuzzy detection | 74.5 (↓) | 56.0 |
| w/o knowledge retrieval and alignment | 56.5 (↓) | 37.4 |
| w/o revision | 72.4 (↓) | 51.7 |

*Note:* Bold values represent the performance value of the comparison between TCSR-SQL and other baseline methods for the main experiment and the ablation experiment.

**TABLE 4** | Main results.

| Methods | EX | EM | Token cost | Average SQL length | Number of joins |
|---|---|---|---|---|---|
| TCSR-SQL | **77.8** | **53.5** | 607 | 113.28 | 0.36 |
| C3-SQL | 30.1 | 22.4 | 1080 | 110.01 | 0.25 |
| DIN-SQL | 32.7 | 28.6 | 10,304 | 118.55 | 0.40 |
| DAIL-SQL | 33.9 | 32.6 | 526 | 107.66 | 0.26 |
| MAC-SQL | 44.2 | 17.7 | 3215 | 116.05 | 0.31 |
| CodeS-15b-spider | 50.0 | 46.7 | **254** | 115.28 | 0.35 |
| CHESS | 33.0 | 21.0 | 2681 | 111.57 | 0.38 |

*Note:* Bold values represent the performance value of the comparison between TCSR-SQL and other baseline methods for the main experiment and the ablation experiment.

retrieving encoding knowledge. Without this module, the EX and EM drops by 21.3% and 16.1%, respectively. The encoding knowledge table stores knowledge that cannot be inferred solely through natural language questions, database schema or data contents. Encoding knowledge is necessary in the real world, especially in practical business. The ability to use encoding knowledge is an important condition for LLM-based methods to replace human labour in text-to-SQL tasks. The lack of this module poses a challenge for TCSR-SQL in determining conditional values, thus proving its necessity.

*w/o Revision*. After generating SQL, no correction is made through execution feedback and the results are directly output.

The SQL revision module executes the generated SQL on the database instance and corrects SQL that cannot be executed or returns empty results, whereas SQL that can be executed to the result does not need to be corrected. Without this module, the EX and EM decreases by 5.4% and 1.8%, respectively. The execution results of SQL are divided into three types: 1. can be executed normally and obtain results, 2. return empty results after execution and 3. cannot be executed normally and report errors. This module collects error or empty result information, feeding it back to LLM for correction until the SQL executes successfully and yields a result or the maximum attempt limit is reached. Since SQL which cannot be executed or has no execution result is wrong, the SQL Revision module can help correct these SQL and improve the performance.

## 4.4 | Robustness to General Benchmarks

To verify the robustness of TCSR-SQL on the text-to-SQL general benchmark, we compare DIN-SQL and MAC-SQL combined with TCSR-SQL on the Spider-dev set [17]. Specifically, we combine them with TCSR-SQL by replacing the *Fuzzy SQL* with DIN-SQL's official GPT-4 results and MAC-SQL's reproduced GPT-4-turbo results. It should be noted that BIRD [18] is not selected here because it provides the evidence for each question-SQL pair, including the exact stored database value corresponding to the data content keywords for SQL generation. Therefore, using BIRD cannot prove the robustness of TCSR-SQL. Meanwhile, it is unfeasible to achieve this in real-world applications with large databases.

Figure 10 demonstrates that the EX of DIN-SQL and MAC-SQL are enhanced by 2.08% and 0.77% upon integration of TCSR-SQL. These results substantiate the robustness of TCSR-SQL to general text-to-SQL benchmarks because TCSR-SQL can enhance the performance of existing methods when confronted with table content-aware questions.

## 4.5 | Case Study

To better understand how the three modules in TCSR-SQL help LLM generate SQL that can be executed correctly, we demonstrate two example cases selected from the dataset. From Figure 11, we can see that previous methods cannot obtain the correct data content values as query conditions because they cannot check the database content or utilise encoding knowledge. In the first case, the keywords related to the data values in the question are not aligned with the values stored in the database. MAC-SQL infers some data values through data samples, but other methods cannot infer the conditional data values of the SQL, causing the empty execution results of the generated SQL. In the second case, the condition value mentioned in the question also differs from the value stored in the database. Since a small number of data samples cannot cover this data value, MAC-SQL also directly utilises the keyword within the question as the conditional data value. In addition, the column make where the conditional value is located is not mentioned in the question, so relying on LLM's implicit inference leads to the wrong column Model.

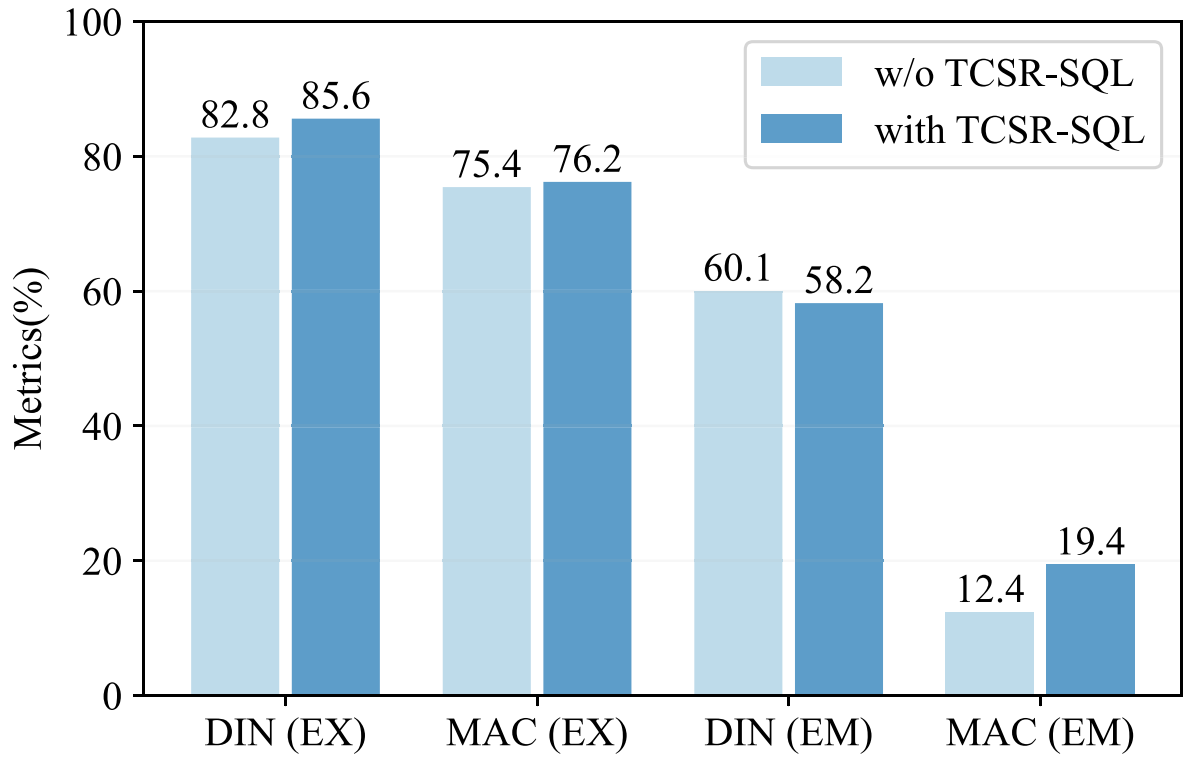

**FIGURE 10** | Robustness of TCSR-SQL on Spider-dev set.

When applying TCSR-SQL in case 1, as shown in Figure 12, the keywords extraction and schema detection module will extract data content keywords and database schema keywords. Specifically, this module uses LLM to infer the tables and columns corresponding to relevant keywords and checks the data contents in the table to obtain exact stored data values. For keywords that cannot obtain stored data values, the knowledge retrieval and alignment module retrieves its corresponding encoding knowledge in the encoding knowledge table. By utilising the encoding knowledge obtained from the above two modules, LLM generates SQL query conditions correctly.

Similarly in case 2, as shown in Figure 13, LLM uses the encoding knowledge obtained by the two modules to generate SQL. The difference is that the condition of the generated SQL query is still wrong. The SQL generation and revision module obtains feedback that the result is empty by executing the SQL and uses it to regenerate SQL that can be executed correctly.

## 4.6 | Discussion

### 4.6.1 | Effectiveness of TCSR-SQL

As shown in Table 4, TCSR-SQL achieves at least a 27.8% improvement in execution accuracy on the TCD dataset compared to existing state-of-the-art (SOTA) methods. This improvement is primarily attributed to the two self-retrieval techniques in TCSR-SQL, which enable effective retrieval of the database schema column names and exact content values.

| Question 1 | How many inflows were there on Jan 31 in Chengjiao, Conghua, Guangzhou? ( 1月31日在广州从化的城郊这里有多少流入人口?) | result | |
|---|---|---|---|
| Gold SQL | SELECT liuru_cnt FROM population where city_nam = "Guangzhou city" ("广州市") and dist_nam = "Conghua district" ("从化区") and street_nm = "Chengjiao street" ("城郊街道") and int_day = "20240131"; | 85952 | |
| C3-SQL | SELECT int_day, liuru_cnt FROM population WHERE city_nam = 'Guangzhou' ('广州') AND dist_nam = 'Conghua' ('从化') AND int_day = 'Jan 31' ('1月31日'); | None | ✗ |
| DIN-SQL | SELECT liuru_cnt FROM population WHERE prov_nam = 'Guangzhou' ('广州') AND city_nam = 'Conghua' ('从化') AND dist_nam = 'Chengjiao' ('城郊') AND int_day = 'Jan 31' ('1月31日'); | None | ✗ |
| DAIL-SQL | SELECT SUM(liuru_cnt) FROM population WHERE city_nam = "Guangzhou"("广州") AND dist_nam = "Conghua"("从化") AND int_day = 31; | None | ✗ |
| MAC-SQL | SELECT liuru_cnt FROM population WHERE city_nam = 'Guangzhou' ('广州') AND dist_nam = 'Conghua' ('从化') AND street_nm = 'Chengjiao' ('城郊') AND int_day = 20240131 | None | ✗ |
| CodeS-15B-spider | SELECT changzhu_cnt FROM population WHERE int_day = "20240131" AND city_nam = "Guangzhou city" ("广州市") AND dist_nam = "Conghua district" ("从化区") ; | 11483, 71236, … | ✗ |
| CHESS | SELECT SUM(liuru_cnt) FROM population WHERE city_nam = 'Guangzhou' ('广州') AND dist_nam = 'Conghua' ('从化') AND int_day = 31; | None | ✗ |
| TCSR-SQL | SELECT population.liuru_cnt FROM population WHERE int_day = 20240131 AND city_nam = "Guangzhou city" ("广州市") AND dist_nam = "Conghua district" ("从化区") AND street_nm = "Chengjiao street" ("城郊街道") | 85952 | ✓ |
| **Question 2** | **What's the rate acceleration of 'amc hornet sports about sw'?** | **result** | |
| Gold SQL | SELECT T1.Accelerate FROM CARS_DATA AS T1 JOIN CAR_NAMES AS T2 ON T1.Id = T2.MakeId WHERE T2.Make = 'amc hornet sportabout (sw)' | 13.5 | |
| C3-SQL | SELECT accelerate FROM cars_data JOIN car_names ON cars_data.id = car_names.makeid JOIN model_list ON car_names.model = model_list.model JOIN car_makers ON model_list.maker = car_makers.id WHERE car_names.model = 'amc hornet sports about sw'; | None | ✗ |
| DIN-SQL | SELECT Accelerate FROM cars_data JOIN car_names ON cars_data.Id = car_names.MakeId JOIN model_list ON car_names.Model = model_list.Model WHERE model_list.Model = 'amc hornet sports about sw' | None | ✗ |
| DAIL-SQL | SELECT Accelerate FROM cars_data WHERE Id = (SELECT MakeId FROM car_names WHERE Model = 'amc hornet sports about sw') | None | ✗ |
| MAC-SQL | SELECT Accelerate FROM cars_data WHERE Id = (SELECT MakeId FROM car_names WHERE Model = 'amc hornet sports about sw') | None | ✗ |
| CodeS-15B-spider | SELECT accelerate FROM cars_data JOIN car_names ON cars_data.id = car_names.makeid WHERE car_names.model = "amc hornet sports about sw" | None | ✗ |
| CHESS | SELECT cars_data.Accelerate FROM cars_data JOIN model_list ON cars_data.Id = model_list.ModelId WHERE model_list.Model = 'amc hornet sports about sw'; | None | ✗ |
| TCSR-SQL | SELECT cars_data.Accelerate FROM cars_data JOIN car_names ON cars_data.Id = car_names.MakeId WHERE car_names.Make = "amc hornet sportabout (sw)" | 13.5 | ✓ |

**FIGURE 11** | Case study on baselines and TCSR-SQL on two questions.

| Question: | How many inflows were there on Jan 31 in Chengjiao, Conghua, Guangzhou? (1月31日在广州从化的城郊这里有多少流入人口？) | | | |
|---|---|---|---|---|
| Keywords Extraction & Schema Detection | Data content keyword "Jan 31" ("1月31日") | → | population.int_day | ✓ |
| | Data content keyword "Guangzhou" ("广州") | → | population.city_nam "Guangzhou city" ("广州市") | ✓ |
| | Data content keyword "Conghua" ("从化") | → | population.city_nam | ✗ |
| | Data content keyword "Chengjiao"("城郊") | → | population.dist_nam | ✗ |
| | Schema keyword "inflows" ("流入人口") | → | population.liuru_cnt | ✓ |
| Knowledge Retrieval & Alignment | Data content keyword "Conghua" ("从化") | → | population. dist_nam "Conghua district" ("从化区") | ✓ |
| | Data content keyword "Chengjiao" ("城郊") | → | population.street_nm "Chengjiao district" ("城郊街道") | ✓ |
| SQL Generation & Revison | **SQL generated:** SELECT population.liuru_cnt FROM population WHERE int_day = 20240131 AND city_nam = "Guangzhou city" ("广州市") AND dist_nam = "Conghua district" ("从化区") AND street_nm = "Chengjiao district" ("城郊街道") | | **Execution result:** 85952 | ✓ |
| | No need to revision | | | |

**FIGURE 12** | Case study of Question 1 on TCSR-SQL.

| Question: | What's the rate acceleration of 'amc hornet sports about sw'? | | | |
|---|---|---|---|---|
| Keywords Extraction & Schema Detection | Data content keyword "amc hornet sportabout" | → | model_list.Model | ✗ |
| | Schema keyword "accelerate" | → | cars_data.Accelerate | ✓ |
| Knowledge Retrieval & Alignment | Data content keyword "amc hornet sportabout" | → | car_names.Make "amc hornet sportabout (sw)" | ✓ |
| SQL Generation & Revison | **SQL generated:** SELECT Accelerate FROM cars_data JOIN car_names ON cars_data.Id = car_names.MakeId WHERE Model = "amc hornet sportabout (sw)" | | **Execution result:** None | ✗ |
| | **SQL revised:** SELECT cars_data.Accelerate FROM cars_data JOIN car_names ON cars_data.Id = car_names.MakeId WHERE car_names.Make = "amc hornet sportabout (sw)" | | **Execution result:** 13.5 | ✓ |

**FIGURE 13** | Case study of Question 2 on TCSR-SQL.

For example, Figure 11 illustrates that existing SOTA methods struggle to leverage LLMs to accurately access database contents. They either fail to retrieve accurate data content values as SQL conditions or miss essential columns used in the SQL query. In contrast, Figures 12 and 13 show that the knowledge retrieval and alignment module in TCSR-SQL guides the LLM to effectively retrieve relevant database information according to the keywords extracted by the keywords extraction and schema detection module. Through prompt engineering and in-context learning, LLM utilises this information efficiently during SQL generation and revision. Moreover, Figure 10 demonstrates that TCSR-SQL enhances existing SOTA methods by enabling accurate self-retrieval of column names and exact content values stored in the database, leading to improved performance on the general text-to-SQL benchmark (Spider-dev set). These results collectively verify the effectiveness of TCSR-SQL in generating correct SQL for table content-aware questions.

#### 4.6.2 | Limitation

The limitation of TCSR-SQL lies in the semi-automated construction of the encoding knowledge table. The current approach requires manual collection of static knowledge and automated acquisition of dynamic knowledge based on input questions. Figure 14 shows that TCSR-SQL is sensitive to the encoding knowledge table since removing different proportions of data from this table results in a decline in execution accuracy ranging from 3.9% to 21.3%. Future work may explore the automatic construction of the encoding knowledge table, such as using information retrieval to automatically link data content keywords to the corresponding database schema, thus building the table dynamically.

## 5 | Conclusion

In this paper, we propose towards table content-aware text-to-SQL with self-retrieval (TCSR-SQL). TCSR-SQL utilises self-retrieval to generate SQL queries with both database exact stored content values and corresponding database schema column names, which solves the widespread problem encountered in real-world scenarios. The keywords extraction and fuzzy detection module infers possible related columns and relevant content values stored in the database corresponding to the question. The knowledge retrieval and alignment module identifies the exact column names and content values stored in the database used in the SQL queries. The SQL generation and revision module generates the final SQL query according to the multi-turn execution feedback of each SQL revision result. Extensive experimental results show the superior performance of TCSR-SQL compared to other SOTA LLM-based text-to-SQL methods in terms of two metrics. Future work will involve the automated generation of the encoding knowledge table and the application of different LLMs to specific modules.

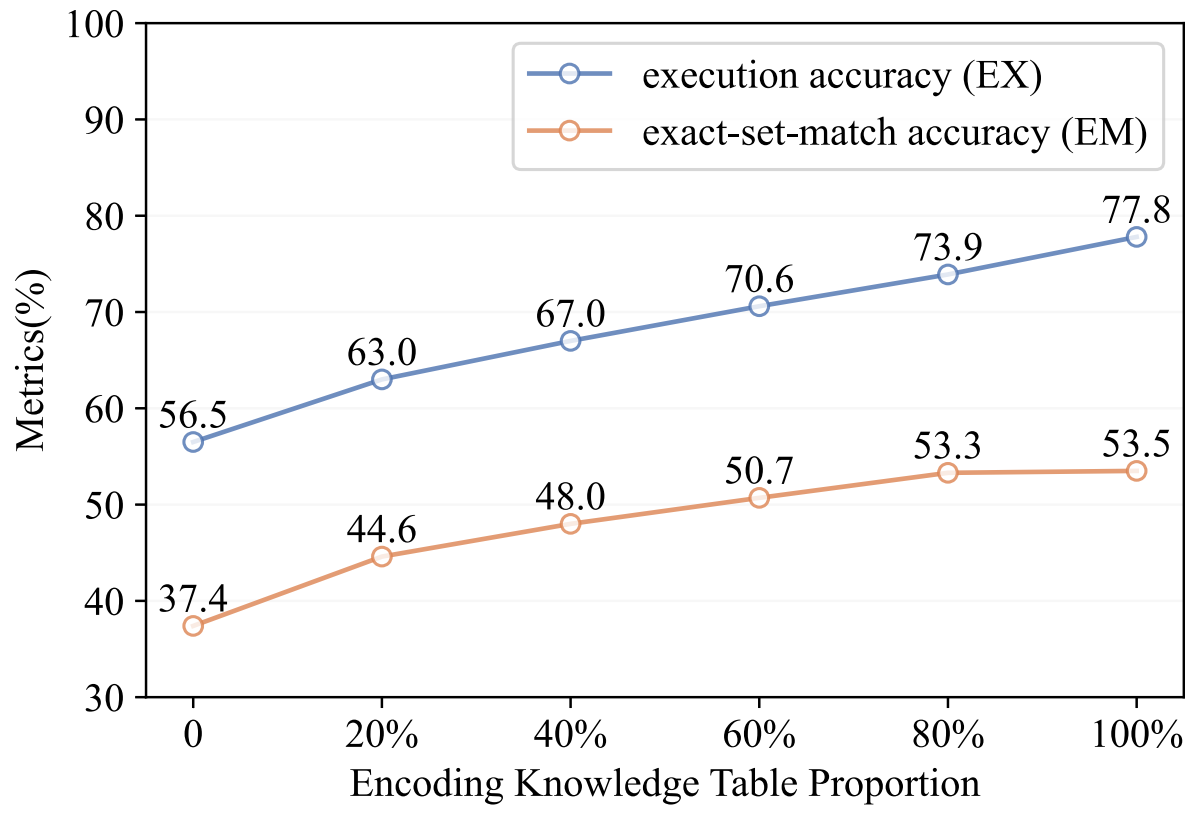

**FIGURE 14** | Sensitivity towards the encoding knowledge table.

### Acknowledgements

This study is supported by the National Key Research and Development Program of China under (Grant 2023YFB3106504), Guangdong Provincial Key Laboratory of Novel Security Intelligence Technologies under (Grant 2022B1212010005), the Major Key Project of PCL under (Grant PCL2023A09), Shenzhen Science and Technology Program under (Grants ZDSYS20210623091809029 and RCBS20221008093131089) and the project of Guangdong Power Grid Co. Ltd. under (Grants 037800KC23090005 and GD-KJXM20231042).

### Conflicts of Interest

The authors declare no conflicts of interest.

### Data Availability Statement

The data that support the findings of this study are available from the corresponding author upon reasonable request.

### Endnotes

[1] https://huggingface.co/DMetaSoul/Dmeta-embedding.

### References

1. V. Zhong, C. Xiong, and R. Socher, "Seq2sql: Generating Structured Queries From Natural Language Using Reinforcement Learning," preprint arXiv:1709.00103 (2017).

2. S. Wu, O. Irsoy, S. Lu, et al., "Bloomberggpt: A Large Language Model for Finance," preprint arXiv:2303.17564 (2023).

3. T. Han, L. C. Adams, J.-M. Papaioannou, et al., "Medalpaca–an Open-Source Collection of Medical Conversational Ai Models and Training Data," preprint arXiv:2304.08247 (2023).

4. Z. Zheng, Y. Deng, D. Xue, Y. Zhou, F. Ye, and Q. Gu, "Structure-Informed Language Models Are Protein Designers," in *International Conference on Machine Learning* (PMLR, 2023), 42317–42338.

5. C. Shen, L. Cheng, Y. You, and L. Bing, "Are Large Language Models Good Evaluators for Abstractive Summarization?," preprint arXiv: 2305.13091 (2023).

6. L. Cheng, X. Li, and L. Bing, "Is GPT-4 a Good Data Analyst?," preprint arXiv:2305.15038 (2023).

7. X. Dong, C. Zhang, Y. Ge, et al., "C3: Zero-Shot Text-to-SQL With Chatgpt," preprint arXiv:2307.07306 (2023).

8. M. Pourreza and D. Rafiei, "Din-SQL: Decomposed in-Context Learning of Text-to-SQL With Self-Correction," *Advances in Neural Information Processing Systems* 36 (2024): 36339–36348, https://doi.org/10.48550/arXiv.2304.11015.

9. D. Gao, H. Wang, Y. Li, et al., "Text-To-SQL Empowered by Large Language Models: A Benchmark Evaluation," preprint arXiv:2308.15363 (2023).

10. C. Zhang, Y. Mao, Y. Fan, et al., "Finsql: Model-Agnostic Llms-Based Text-to-SQL Framework for Financial Analysis," preprint arXiv: 2401.10506 (2024): 93–105, https://doi.org/10.1145/3626246.3653375.

11. B. Wang, C. Ren, J. Yang, et al., "Mac-Sql: Multi-Agent Collaboration for Text-to-SQL," preprint arXiv:2312.11242 (2023).

12. H. Li, J. Zhang, H. Liu, et al., "Codes: Towards Building Open-Source Language Models for Text-to-SQL," preprint arXiv:2402.16347 2, no. 3 (2024): 1–28, https://doi.org/10.1145/3654930.

13. S. Talaei, M. Pourreza, Y.-C. Chang, A. Mirhoseini, and A. Saberi, "Chess: Contextual Harnessing for Efficient Sql Synthesis," preprint arXiv:2405.16755 (2024).

14. S. Patnaik, H. Changwal, M. Aggarwal, S. Bhatia, Y. Kumar, and B. Krishnamurthy, "Cabinet: Content Relevance Based Noise Reduction for Table Question Answering," preprint arXiv:2402.01155 (2024).

15. V. Kumar, Y. Gupta, S. Chemmengath, et al., "Multi-Row, Multi-Span Distant Supervision for Table+ Text Question Answering," in *Proceedings of the 61st Annual Meeting of the Association for Computational Linguistics*, Vol. 1 (Long Papers), 2023), 8080–8094, https://doi.org/10.18653/v1/2023.acl-long.449.

16. P. Lewis, E. Perez, A. Piktus, et al., "Retrieval-Augmented Generation for Knowledge-Intensive Nlp Tasks," *Advances in Neural Information Processing Systems* 33 (2020): 9459–9474, https://doi.org/10.48550/arXiv.2005.11401.

17. T. Yu, R. Zhang, K. Yang, et al., "Spider: A Large-Scale Human-Labeled Dataset for Complex and Cross-Domain Semantic Parsing and Text-to-SQL Task," preprint arXiv:1809.08887 (2018).

18. J. Li, B. Hui, G. Qu, et al., "Can Llm Already Serve as a Database Interface? a Big Bench for Large-Scale Database Grounded Text-to-Sqls," *Advances in Neural Information Processing Systems* 36 (2024), https://doi.org/10.48550/arXiv.2305.03111.

19. H. Li, J. Zhang, C. Li, and H. Chen, "Resdsql: Decoupling Schema Linking and Skeleton Parsing for Text-to-Sql," in *Proceedings of the AAAI Conference on Artificial Intelligence* 37, no. 11, 2023, 13067–13075, https://doi.org/10.1609/aaai.v37i11.26535.

20. H. Zhang, R. Cao, L. Chen, H. Xu, and K. Yu, "Act-Sql: In-Context Learning for Text-to-Sql With Automatically-Generated Chain-of-Thought," preprint arXiv:2310.17342 (2023): 3501–3532, https://doi.org/10.18653/v1/2023.findings-emnlp.227.

21. W. Zhai, X. Zhao, J. Liao, and Z. Chen, "Learn From Mistakes: Guidance on Zero-Shot Conversational Text-to-Sql," in *Proceedings of the 33rd ACM International Conference on Information and Knowledge Management*, (2024), 4243–4247.

22. Y. Xie, X. Jin, T. Xie, et al., "Decomposition for Enhancing Attention: Improving Llm-Based Text-to-Sql Through Workflow Paradigm," preprint arXiv:2402.10671 (2024): 10796–10816, https://doi.org/10.18653/v1/2024.findings-acl.641.

23. T. Ren, Y. Fan, Z. He, et al., "Purple: Making a Large Language Model a Better Sql Writer," preprint arXiv:2403.20014 (2024): 15–28, https://doi.org/10.1109/icde60146.2024.00009.

24. G. Qu, J. Li, B. Li, et al., "Before Generation, Align It! a Novel and Effective Strategy for Mitigating Hallucinations in Text-to-Sql Generation," preprint arXiv:2405.15307 (2024): 5456–5471, https://doi.org/10.18653/v1/2024.findings-acl.324.

25. R. Sun, S. O. Arik, H. Nakhost, et al., "Sql-Palm: Improved Large Language Modeladaptation for Text-to-Sql," preprint arXiv:2306.00739 (2023).

26. R. Anil, A. M. Dai, O. Firat, et al., "Palm 2 Technical Report," preprint arXiv:2305.10403 (2023).

27. E. J. Hu, Y. Shen, P. Wallis, et al., "Lora: Low-Rank Adaptation of Large Language Models," preprint arXiv:2106.09685 (2021).

28. H. Touvron, L. Martin, K. Stone, et al., "Llama 2: Open Foundation and Fine-Tuned Chat Models," preprint arXiv:2307.09288 (2023).

29. A. Yang, B. Xiao, B. Wang, et al., "Baichuan 2: Open Large-Scale Language Models," preprint arXiv:2309.10305 (2023).

30. M. Pourreza and D. Rafiei, "Dts-Sql: Decomposed Text-to-Sql With Small Large Language Models," preprint arXiv:2402.01117 (2024): 8212–8220, https://doi.org/10.18653/v1/2024.findings-emnlp.481.

31. R. Sun, S. Ö. Arik, R. Sinha, et al., "Sqlprompt: In-Context Text-to-Sql With Minimal Labeled Data," preprint arXiv:2311.02883 (2023): 542–550, https://doi.org/10.18653/v1/2023.findings-emnlp.39.

32. S. Volvovsky, M. Marcassa, and M. Panbiharwala, "Dfin-Sql: Integrating Focused Schema With din-sql for Superior Accuracy in Large-Scale Databases," preprint arXiv:2403.00872 (2024).

33. Z. Li, X. Wang, J. Zhao, et al., "Pet-Sql: A Prompt-Enhanced Two-Stage Text-to-Sql Framework With Cross-Consistency," preprint arXiv:2403.09732 (2024).

34. D. Lee, C. Park, J. Kim, and H. Park, "Mcs-Sql: Leveraging Multiple Prompts and Multiple-Choice Selection for Text-to-Sql Generation," preprint arXiv:2405.07467 (2024).

35. J. Chen, "Sqlcritic: Correcting Text-to-Sql Generation via Clause-Wise Critic," preprint arXiv:2503.07996 (2025).

36. C. Guo, Z. Tian, J. Tang, et al., "Retrieval-Augmented Gpt-3.5-Based Text-to-Sql Framework With Sample-Aware Prompting and Dynamic Revision Chain," in *International Conference on Neural Information Processing* (Springer, 2023), 341–356.

37. H. Xia, F. Jiang, N. Deng, et al., "Sql-Craft: Text-To-Sql Through Interactive Refinement and Enhanced Reasoning," preprint arXiv:2402.14851 (2024).

38. Z. Hong, Z. Yuan, H. Chen, Q. Zhang, F. Huang, and X. Huang, "Knowledge-To-Sql: Enhancing Sql Generation With Data Expert Llm," preprint arXiv:2402.11517 (2024): 10997–11008, https://doi.org/10.18653/v1/2024.findings-acl.653.

39. D. G. Thorpe, A. J. Duberstein, and I. A. Kinsey, "Dubo-Sql: Diverse Retrieval-Augmented Generation and Fine Tuning for Text-to-Sql," preprint arXiv:2404.12560 (2024) [Online], https://api.semanticscholar.org/CorpusID:269282528.

40. R. Liu, S. Yuan, A. Dai, et al., "Few-Shot Table Understanding: A Benchmark Dataset and Pre-Training Baseline," in *Proceedings of the 29th International Conference on Computational Linguistics*, (2022), 3741–3752.

41. A. Askari, C. Poelitz, and X. Tang, "Magic: Generating Self-Correction Guideline for in-Context Text-to-Sql," preprint arXiv: 2406.12692 (2024).

42. J. Cen, J. Liu, Z. Li, and J. Wang, "Sqlfixagent: Towards Semantic-Accurate Sql Generation Via Multi-Agent Collaboration," preprint arXiv:2406.13408 (2024).

43. K. Serebryany, "Continuous Fuzzing With Libfuzzer and Address-sanitizer," in *2016 IEEE Cybersecurity Development (Secdev)* (IEEE, 2016), 157.

44. J. Liu, C. S. Xia, Y. Wang, and L. Zhang, "Is Your Code Generated by Chatgpt Really Correct? Rigorous Evaluation of Large Language Models for Code Generation," *Advances in Neural Information Processing Systems* 36 (2024), https://doi.org/10.48550/arXiv.2305.01210.

45. X. Chen, M. Lin, N. Schärli, and D. Zhou, "Teaching Large Language Models to Self-Debug," preprint arXiv:2304.05128 (2023).

46. Y. Gan, X. Chen, and M. Purver, "Exploring Underexplored Limitations of Cross-Domain Text-to-Sql Generalization," in *Proceedings of*

*the 2021 Conference on Empirical Methods in Natural Language Processing*, (2021), 8926–8931.

47. Y. Gan, X. Chen, Q. Huang, et al., "Towards Robustness of Text-to-Sql Models Against Synonym Substitution," in *Proceedings of the 59th Annual Meeting of the Association for Computational Linguistics and the 11th International Joint Conference on Natural Language Processing*, Vol. 1 (Long Papers, 2021), 2505–2515.

48. X. Deng, A. Hassan, C. Meek, O. Polozov, H. Sun, and M. Richardson, "Structure-Grounded Pretraining for Text-to-Sql," in *Proceedings of the 2021 Conference of the North American Chapter of the Association for Computational Linguistics: Human Language Technologies*, (2021), 1337–1350.

49. S. Chang, J. Wang, M. Dong, et al., "Dr. Spider: A Diagnostic Evaluation Benchmark Towards Text-to-Sql Robustness," in *The Eleventh International Conference on Learning Representations*, (2022).

50. OpenAI. Chatgpt. 2023, https://chat.openai.com/chat.